\documentclass{ws-procs975x65}
\usepackage{graphicx}
\newcommand{\ba} {\begin{eqnarray}}
\newcommand{\ea} {\end{eqnarray}}
\usepackage{dcolumn} 
\usepackage{bm}      

\begin{document}

\title{\hfill {\small\bf ECTP-2009-9}\\
HUBBLE PARAMETER IN BULK VISCOUS COSMOLOGY}

\author{A.~Tawfik$^1\,$, H.~Mansour$^2\,$, M.~Wahba$^1\,$ \\~\\
{\small $^1$ Egyptian Center for Theoretical Physics (ECTP), MTI University,
 Cairo-Egypt} \\
{\small $^2$ Department of Physics, Faculty of Science, Cairo University, Giza-Egypt}
\email{drtawfik@mti.edu.eg}
}

\begin{abstract}

We discuss influences of bulk viscosity on the Early Universe, which is modeled by Friedmann-Robertson-Walker metric and Einstein field equations. We assume that the matter filling the isotropic and homogeneous background is relativistic viscous characterized by ultra-relativistic equations of state deduced from recent lattice QCD simulations. We obtain a set of complicated differential equations, for which we suggest approximate solutions for Hubble parameter $H$. We find that finite viscosity in Eckart and Israel-Stewart fluids would significantly modify our picture about the Early Universe. 

\end{abstract}

\bodymatter

\section{Introduction}\label{sec:intro}

Bulk viscosity (BV) is of essential importance in Early Universe, because it can occur in an isotropic and homogeneous matter. Although, BV likely vanishes at very high temperatures, it remains important in mixed phases of matter. It diverges near the phase transition(s). First-order theory for relativistic fluids has been suggested by Eckart \cite{Ec40} and Landau and Lifshitz \cite{LaLi87}. A relativistic ''second-order'' theory has been introduced by Israel and Stewart (IS) \cite{Is76,IsSt76}. Due to the complicated characters of  resulting differential equations, very few solutions of gravitational field equations are known in framework of causal fluid theory. For technical reasons, most investigations for dissipative causal cosmologies are assuming symmetric Friedmann-Robertson-Walker (FRW) metric.  

Recent RHIC results give a strong indication that hot, dense and viscous matter has been produced~\cite{reff1} in heavy-ion collisions. This matter likely agrees with the Lattice QCD simulations~\cite{karsch07}, which show that BV is not negligible near QCD critical temperature $T_c$~\cite{karsch07,mueller2}. It has been shown that BV at high temperature $T$ and weak coupling $\alpha_s$, $\xi\sim  \alpha_s^2 T^3/\ln \alpha_s^{-1}$~\cite{mueller3}. Following the prescription in Ref.~\cite{LaLi87}, causal viscous fluid is given by $\xi \propto \rho$, where $\rho $ is energy density.

In this letter, we assume finite bulk viscosity coefficient $\xi$ in Eckert- and IS-fluids and apply QCD EoS to solve Freidmann's equation. We suggest numerical and approximate solutions for Hubble parameter $H$ in dependence on comoving time $t$. We compare our results with the non-viscous fluid.  Also We imply our assumptions to relate $t$ with the temperature $T$.

\section{Model and Results} \label{sec:model}

Using natural units, $k=c=1$ and $h=2\pi$, we assume that background geometry of the Early Universe is filled with bulk viscous fluid, which is given by a spatially flat FRW metric. In spherical coordinates, the line element 
\begin{equation}  \label{1}
ds^{2}=dt^{2}-a^{2}(t) \left[dr^{2}+r^{2}\left( d\theta
^{2}+\sin ^{2}\theta d\phi^{2}\right) \right].
\end{equation}
At vanishing cosmological constant $\Lambda$, Einstein gravitational field equations in a flat Universe ($k=0$) read
\begin{equation}  \label{ein}
R_{ik}-\frac{1}{2}g_{ik}R=8\pi\; G\; T_{ik}. 
\end{equation}
Inclusion of bulk viscous effects can be generalized through an effective
pressure $\Pi$, which is formally included in the effective thermodynamic pressure $p_{eff}$~\cite{Ma95}, where $p_{eff}=p+\Pi $ and $p$ is the thermodynamic pressure. Then, in comoving frame, the energy momentum tensor has the components $T_{0}^{0}=\rho ,T_{1}^{1}=T_{2}^{2}=T_{3}^{3}=-p_{eff}$.
For FRW line element, Eq.~(\ref{1}), field equations become
\begin{equation}  \label{2}
3H^2=\rho,
\hspace*{2cm}
2 \dot H + 3 H^2 = p_{eff},
\end{equation}
where one dot denotes differentiation with respect to $t$. In a closed system, total energy density of cosmic matter is conserved,  i.e., $T_{i;j}^j=0$ and
\begin{equation}  \label{5}
\dot{\rho}+3H\left( p_{eff}+\rho \right) =0,
\end{equation} 
where $H=\dot{a}/a$ is the Hubble parameter. 
In presence of bulk viscous stress $\Pi$, Eq.~(\ref{5}) can be written as
\begin{equation}  \label{6}
\dot{\rho}+3H\left( p+\rho \right) =-3\Pi H.
\end{equation}
According to Eckert theory, bulk viscous stress is given as 
\begin{eqnarray} \label{eckert-pii}
\Pi &=& - 3 \xi H, 
\end{eqnarray}
which has in IS-theory~\cite{Is76,IsSt76}, following evolution:
\begin{equation}  \label{8}
\tau \dot{\Pi}+\Pi =-3\xi H-\frac{1}{2}\tau \Pi \left( 3H+\frac{\dot{\tau}}{%
\tau }-\frac{\dot{\xi}}{\xi }-\frac{\dot{T}}{T}\right) ,
\end{equation}
where $T$ is temperature and $\tau$ is relaxation time.

\begin{figure}
\begin{center}
\includegraphics[width=6cm,angle=0]{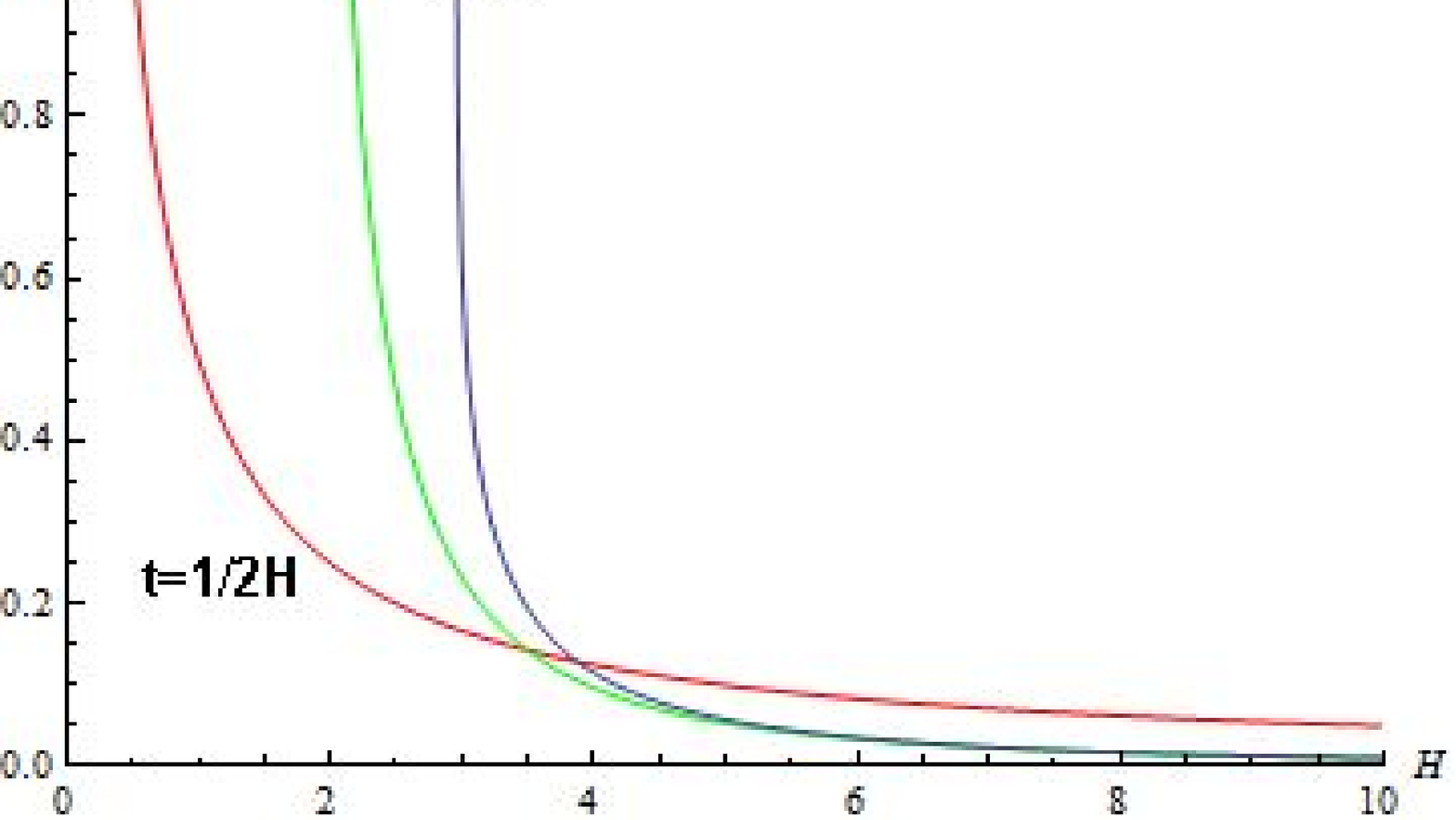} 
\includegraphics[width=6cm,angle=0]{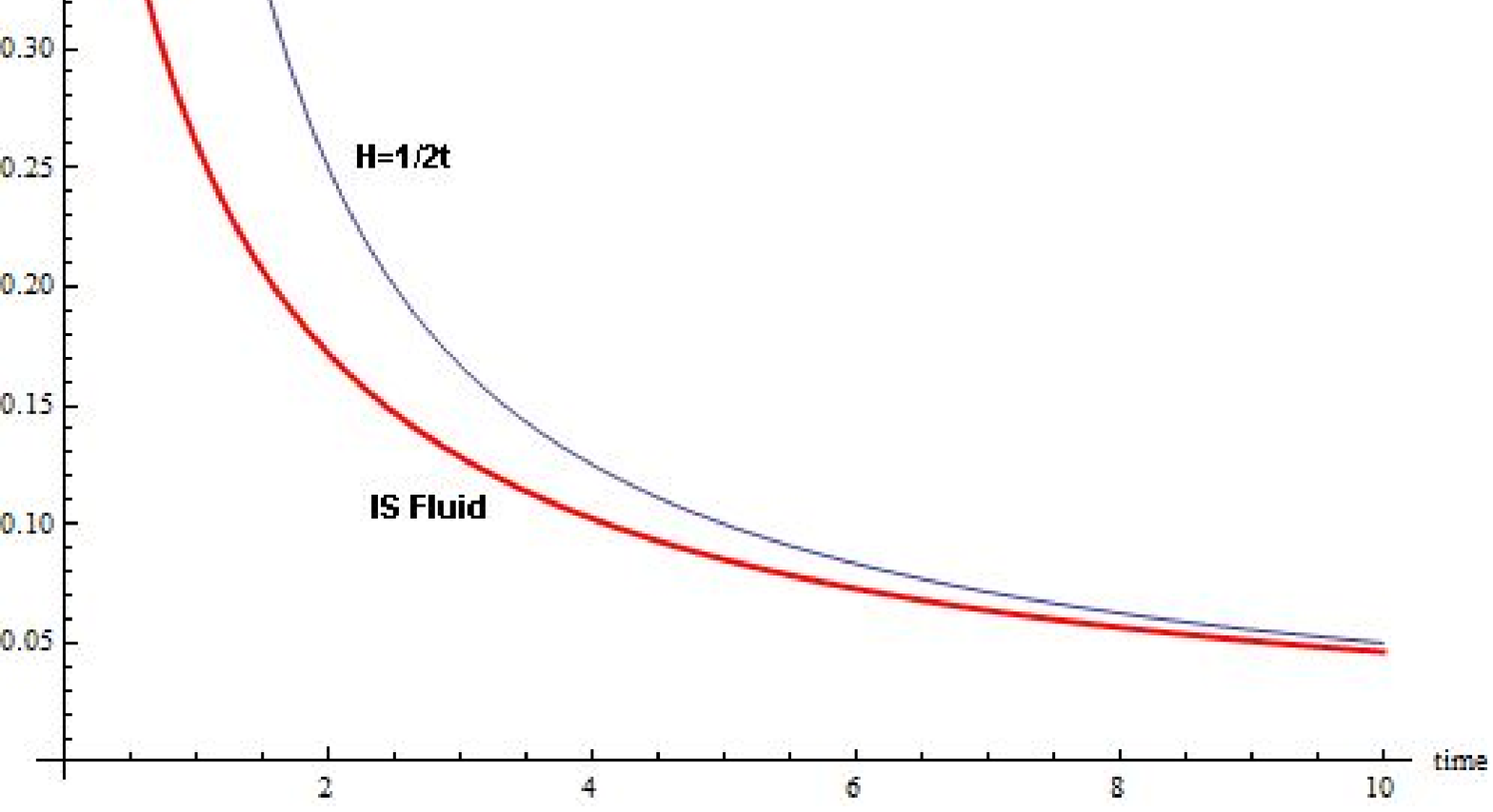}\vspace*{-4.5cm}
\caption{Left panel: $t$ vs. $H(t)$ in Eckert fluid, Eq.~\ref{eckert-tH}, compared with the case when $b$ in Eq.~\ref{13} vanishes and with non-viscous fluid, $t=1/2H$. Right panel: numerical solution of $H(t)$ in IS-fluid, Eq.~\ref{init},  compared with non-viscous fluid.}
\label{fig1}
\end{center}
\end{figure}

EoS for $p$ and $T$ can help to have a closed system from Eq.~(\ref{2}), (\ref{6}) and (\ref{8}). $\tau$ and $\xi$ can be  determined according to phenomenological approaches. For instance, BV in quark-gluon plasma (QGP) at high $T$ can be estimated from recent lattice QCD simulations~\cite{karsch07,Cheng:2007jq}.
\begin{equation} \label{13}
p=\omega \rho, \hspace*{1cm} T=\beta\rho^r, \hspace*{1cm} \xi = \alpha \rho + \frac{9}{\omega_0} T_c^4\equiv \alpha \rho + b, 
\end{equation}
where  $\omega = (\gamma-1)$, $\omega_0 \simeq 0.5-1.5$ $GeV$, 
$\alpha = -(9\gamma^2-24\gamma+16)/[9\omega_0(\gamma-1)]$, $\beta = 0.718$ and 
$\gamma \simeq 1.318$. 
For Eckert fluid, we insert Eq.~\ref{eckert-pii} in Eq.~\ref{6} to get
\begin{eqnarray} \label{eckert-tH}
t &=& \frac{-1}{3b} \left(\frac{2\gamma\arctan\left(\frac{6\alpha H-\gamma}{\sqrt{12\alpha b-\gamma^2}}\right)}{\sqrt{12\alpha b-\gamma^2}} + 2\ln(H) - \ln\left(b-\gamma H+3\alpha H^2\right)\right).
\end{eqnarray}
The results are depicted in left panel of Fig.~\ref{fig1} and compared with the non-viscous fluid, $t=1/2H$. At b=0, i.e., $\xi\propto \rho$, Eq.~\ref{eckert-tH} turns to be
\begin{equation}
t(H)= 2\left[\frac{\gamma-3\alpha H \ln H+3\alpha H \ln(3\alpha H-\gamma)}{3\gamma^2 H}\right].
\end{equation}

For IS fluid, let us first assume that the last term in rhs of Eq.~\ref{8} is too small compared to first term. With this assumption Eq.~\ref{6} can be solved and we get $T\propto R^3/\rho$, which has been discussed in Ref.~\cite{Ma95} as one of unphysical cases, in which $T$ increases with the Universe expansion. Substituting Eq.~\ref{8} into Eq.~\ref{6} leads to 
\begin{eqnarray}\label{init}
\ddot H + \frac{3}{2} [1+(1-r) \gamma] H\dot H + \frac{1}{\alpha}\dot H - 
(1+r) \frac{\dot H^2}{H} + \frac{9}{4}(\gamma -2) H^3 + \frac{3}{2}\frac{\gamma}{\alpha} H^2 = 0
\end{eqnarray}
which is a non-homogeneous and non-linear second type Abel differential equation~\cite{tawfik091}. If we would assume that the last two term in lhs are vanishing, we then left with Bernoulli equation with the solutions $H=0$ and $H\approx -2\gamma/[3\alpha(2-\gamma)]$. In both cases, $H$ does't depend on $t$, i.e., static Universe. Again, if it is possible to express $\xi$ depending on $t$, we get $2\dot H + 3\gamma H^2 -3\xi(t) H =0$, 
which is Bernoulli equation and has the solution, $H(t)=\exp{\left[\frac{3}{2}\int\xi(t)dt\right]}/\left(c+\exp{\left[\frac{3}{2}\gamma \int\xi(t)dt\right]}\right)$. The numerical solution for Eq.~\ref{init} is given in right panel of Fig.~\ref{fig1}. It strongly depends on the initial conditions for $H$ and $\dot H$. \\

\begin{figure}
\begin{center}
\includegraphics[width=8cm,angle=0]{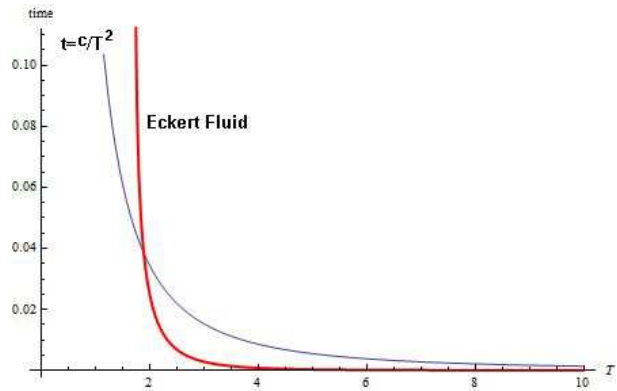} \vspace*{-6cm}
\caption{$t$ vs. $T$ in Eckert fluid, Eq.~\ref{eckert-Ttt}, compared with the non-viscous fluid, $t\propto T^{-2}$.}
\label{fig2}
\end{center}
\end{figure}

One of the direct implications of our assumptions is the relation $t$ vs. $T$. From Eckert solution, $\rho(t)=3H^2(t)$, and Eq.~\ref{13} we get  
\begin{eqnarray}\label{eckert-Ttt}
t&=&\frac{\ln T^{4D}-\ln(-B-D+2AT^2)^{D-B}-\ln(-B+D+2AT^2)^{B+D}}{3CD}
\end{eqnarray}
where $A=3\alpha\beta$, $B=3\gamma(\alpha/3)^{-1/2}$, $C=9T_c^4$, $D=(B^2+12AC)^{1/2}$. Eq.~\ref{eckert-Ttt} is depicted in Fig.~\ref{fig2} and compared with the non-viscous fluid, $t=c T^{-2}$. 

\vspace*{-1.cm}
%


\end{document}